\documentclass[english]{article}
\usepackage[T1]{fontenc}
\usepackage[latin9]{inputenc}
\usepackage{color}
\usepackage{textcomp}
\usepackage{amstext}
\usepackage{amssymb}
\usepackage{esint}

\makeatletter

\newcommand{\lyxmathsym}[1]{\ifmmode\begingroup\def\b@ld{bold}
  \text{\ifx\math@version\b@ld\bfseries\fi#1}\endgroup\else#1\fi}

\makeatother

\usepackage{babel}
\begin{document}

\title{\textbf{Quasi-normal modes: the ``electrons'' of black holes as
``gravitational atoms''? Implications for the black hole information
puzzle}}

\author{\textbf{Christian Corda}}

\maketitle
\begin{center}
Dipartimento di Scienze, Sezione di Fisica, Scuola Superiore di Studi
Universitari e Ricerca \textquotedbl{}Santa Rita\textquotedbl{}, via
San Nicola snc, 81049, San Pietro Infine (CE) Italy
\par\end{center}

\begin{center}
Austro-Ukrainian Institute for Science and Technology, Institut for
Theoretish Wiedner Hauptstrasse 8-10/136, A-1040, Wien, Austria
\par\end{center}

\begin{center}
International Institute for Applicable Mathematics \& Information
Sciences (IIAMIS),  B.M. Birla Science Centre, Adarsh Nagar, Hyderabad
- 500 463, India 
\par\end{center}

\begin{center}
\textit{E-mail address:} \textcolor{blue}{cordac.galilei@gmail.com} 
\par\end{center}
\begin{abstract}
Some recent important results on black hole (BH) quantum physics concerning
the \emph{BH effective state} and the natural correspondence between
Hawking radiation and BH quasi-normal modes (QNMs) are reviewed, clarified
and refined. Such a correspondence permits to naturally interpret
QNMs as quantum levels in a semi-classical model. This is a model
of BH somewhat similar to the historical semi-classical model of the
structure of a hydrogen atom introduced by Bohr in 1913. In a certain
sense, QNMs represent the \textquotedbl{}electron\textquotedbl{} which
jumps from a level to another one and the absolute values of the QNMs
frequencies ``triggered'' by emissions (Hawking radiation) and absorption
of particles represent the energy \textquotedbl{}shells\textquotedbl{}
of the ``gravitational hydrogen atom''. Important consequences on
the BH information puzzle are discussed. In fact, it is shown that
the time evolution of this ``Bohr-like BH model'' obeys to a \emph{time
dependent Schrödinger equation} which permits the final BH state to
be a \emph{pure} quantum state instead of a mixed one. Thus, information
comes out in BH evaporation, in agreement with the assumption by 't
Hooft that Schröedinger equations can be used universally for all
dynamics in the universe. We also show that, in addition, our approach
solves the entanglement problem connected with the information paradox.

We emphasize that Bohr model is an approximated model of the hydrogen
atom with respect to the valence shell atom model of full quantum
mechanics. In the same way, we expect the Bohr-like BH model to be
an approximated model with respect to the definitive, but at the present
time unknown, BH model arising from a full quantum gravity theory.
\end{abstract}

\section{Introduction}

An intriguing and largely used framework to obtain Hawking radiation
\cite{key-1} is the tunnelling mechanism, see \cite{key-2}-\cite{key-8}
for example. Considering an object classically stable, if it becomes
unstable in a quantum-mechanically approach, it is natural to suspect
that a tunnelling process works. The famous Hawking's mechanism of
particles creation by BHs \cite{key-1}, is described in a modern
way as tunnelling of particles near the BH horizon \cite{key-2}-\cite{key-8}.
Let us assume that a virtual particle pair is created just inside
the horizon. Then, the virtual particle having positive energy can
tunnel out and materializes outside the BH as a real particle. The
same happens when one considers a virtual particle pair created just
outside the BH horizon. In that case, the particle having negative
energy can tunnel inwards. The result of both of the situations is
that the BH absorbs the particle having negative energy. Thus, the
BH mass decreases and the particle having positive energy propagates
towards infinity. Subsequent emissions of quanta appear, in turn,
as Hawking radiation. The remarkable result of Parikh and Wilczek
\cite{key-2,key-3} has shown a probability of emission which is compatible
with a non-strictly thermal spectrum, differently from the original
result of Hawking \cite{key-1} and a recent result of Banerjee and
Majhi \cite{key-7}, who found a perfect black body spectrum through
a reformulation of the tunnelling mechanism. In \cite{key-8} we have
recently improved the tunnelling approach in \cite{key-2,key-3},
showing that the obtained probability of emission is really associated
to the two distributions \cite{key-8,key-20} 
\begin{equation}
<n>_{boson}=\frac{1}{\exp\left[4\pi\left(2M-\omega\right)\omega\right]-1},\;\;<n>_{fermion}=\frac{1}{\exp\left[4\pi\left(2M-\omega\right)\omega\right]+1},\label{eq: final distributions}
\end{equation}
for bosons and fermions respectively, which are \emph{non-strictly}
thermal. The derivation of the two distributions (\ref{eq: final distributions})
will be sketched in Section 2 of this paper.

The non precise black body spectrum has important implications for
the BH information puzzle. In fact, arguments that information is
lost during BH evaporation partially rely on the assumption of strictly
black body spectrum \cite{key-3,key-9}. The non-strictly thermal
behavior in \cite{key-2,key-3} implies that emissions of subsequent
Hawking quanta are countable \cite{key-8}, \cite{key-10}-\cite{key-20},
and, in turn, generates a natural correspondence between Hawking radiation
and BH QNMs \cite{key-10}-\cite{key-14} \cite{key-20,key-46}, permitting
to naturally interpret QNMs as quantum levels \cite{key-10}-\cite{key-14},
\cite{key-20,key-46}. The system composed by Hawking radiation and
BH QNMs is somewhat similar to the semi-classical Bohr model of the
structure of a hydrogen atom \cite{key-43,key-44,key-45}. In the
BH model in \cite{key-10}-\cite{key-14}, \cite{key-20,key-46},
during a quantum jump a discrete amount of energy is indeed radiated
and, for large values of the principal quantum number $n,$ the analysis
becomes independent of the other quantum numbers. In a certain sense,
QNMs represent the \textquotedbl{}electron\textquotedbl{} which jumps
from a level to another one and the absolute values of the QNMs frequencies
represent the energy \textquotedbl{}shells\textquotedbl{} \cite{key-12,key-20}.
In Bohr model \cite{key-43,key-44,key-45} electrons can only gain
and lose energy by jumping from one allowed energy shell to another,
absorbing or emitting radiation with an energy difference of the levels
according to the Planck relation $E=hf$, where $\: h\:$ is the Planck
constant and $f\:$ the transition frequency. In the BH model in \cite{key-10}-\cite{key-14},
\cite{key-20,key-46}, QNMs can only gain and lose energy by jumping
from one allowed energy shell to another, absorbing or emitting radiation
(emitted radiation is given by Hawking quanta) with an energy difference
of the levels according to equations which are in full agreement with
previous literature of BH thermodynamics, like references \cite{key-25,key-35,key-36}.
More, it will be shown that the BH model in \cite{key-10}-\cite{key-14},
\cite{key-20,key-46} is also in agreement with the famous result
of Bekenstein on the \emph{area quantization} \cite{key-34}. Bekenstein
\cite{key-54} was also, in our knowledge, the first who viewed BHs
as similar to Bohr atoms \cite{key-53}, although our model in \cite{key-10}-\cite{key-14},
\cite{key-20,key-46} is more detailed than Bekenstein's original
intuition. 

It is important to recall that Bohr model is an approximated model
of the hydrogen atom with respect to the valence shell atom model
of full quantum mechanics. In the same way, one expects the Bohr-like
BH model to be an approximated model with respect to the definitive,
but at the present time unknown, BH model arising from a definitive
theory of quantum gravity. 

The time evolution of the Bohr-like BH model obeys a \emph{time dependent
Schrödinger equation }which permits to write the physical state and
the correspondent\emph{ wave function} in terms of an \emph{unitary}
evolution matrix instead of a density matrix \cite{key-20}. The BH
final state results in turn to be a \emph{pure} quantum state instead
of mixed one \cite{key-20}. The fundamental consequence is that information
comes out in BH evaporation in terms of pure states in an unitary
time dependent evolution \cite{key-20}. Thus, the BH evolution is
in full agreement with the assumption by 't Hooft that Schröedinger
equations can be used universally for all dynamics in the universe
and this endorses the conclusion that BH evaporation is information
preserving \cite{key-20}. In addition, we will see that the present
approach permits also to solve the entanglement problem connected
with the BH information puzzle \cite{key-20}.

\section{Deviation from standard distributions for bosons and fermions}

A problem on the tunnelling approach for Hawking radiation was that,
in \cite{key-4,key-5,key-6} and in other papers in the literature,
the analysis has been finalized almost only to obtain the Hawking
temperature through a comparison of the probability of emission of
an outgoing particle with the Boltzmann factor. In the interesting
work \cite{key-7} the problem was apparently solved. In fact, analysing
the tunnelling mechanism in a slightly different way, the authors
of \cite{key-7} found a perfect black body spectrum for Hawking radiation.
On the other and, the result in \cite{key-7} is in contrast which
the result in \cite{key-2,key-3}, that, indeed, has shown a probability
of emission which is compatible with a non-strictly black body spectrum
having associated two non-strictly thermal distributions. Introducing
a BH effective state \cite{key-8}, \cite{key-10}-\cite{key-14},
\cite{key-20,key-46}, we solved such a contradiction in \cite{key-8},
considering a modification of the analysis in \cite{key-7}. The final
result, which we review in this Section, is a non-strictly black body
spectrum with associated two non-strictly thermal distributions in
BH evaporation.

For the sake of simplicity, in all this paper we work with Planck
units, i.e. $G=c=k_{B}=\hbar=\frac{1}{4\pi\epsilon_{0}}=1$. 

Considering a Schwarzschild BH the Schwarzschild line element is \cite{key-8,key-21}
\begin{equation}
ds^{2}=-(1-\frac{2M}{r})dt^{2}+\frac{dr^{2}}{1-\frac{2M}{r}}+r^{2}(\sin^{2}\theta d\varphi^{2}+d\theta^{2})\label{eq: Hilbert}
\end{equation}
(historical notes to this notion can be found in \cite{key-22}).
The Schwarzschild radius (event horizon) is given by $r_{H}=2M$ \cite{key-7,key-8,key-21}
and $\frac{1}{4M}\:$ is the BH surface gravity. As Hawking radiation
will be discussed like tunnelling, the radial trajectory is the only
relevant \cite{key-2,key-3,key-7,key-8}. Following \cite{key-7},
the (normalized) physical states of the system for bosons and fermions
are \cite{key-7}

\begin{equation}
\begin{array}{c}
|\Psi>_{boson}=\left(1-\exp\left(-8\pi M\omega\right)\right)^{\frac{1}{2}}\sum_{n}\exp\left(-4\pi nM\omega\right)|n_{out}^{(L)}>\otimes|n_{out}^{(R)}>\\
\\
|\Psi>_{fermion}=\left(1+\exp\left(-8\pi M\omega\right)\right)^{-\frac{1}{2}}\sum_{n}\exp\left(-4\pi nM\omega\right)|n_{out}^{(L)}>\otimes|n_{out}^{(R)}>.
\end{array}\label{eq: physical states}
\end{equation}
In the following the analysis will be focused only on bosons. For
fermions the treatment is indeed the same \cite{key-7}. The density
matrix operator of the system is given by \cite{key-7}

\begin{equation}
\begin{array}{c}
\hat{\rho}_{boson}\equiv\Psi>_{boson}<\Psi|_{boson}\\
\\
=\left(1-\exp\left(-8\pi M\omega\right)\right)\sum_{n,m}\exp\left[-4\pi(n+m)M\omega\right]|n_{out}^{(L)}>\otimes|n_{out}^{(R)}><m_{out}^{(R)}|\otimes<m_{out}^{(L)}|.
\end{array}\label{eq: matrice densita}
\end{equation}
One traces out the ingoing modes obtaining the density matrix for
the outgoing (right) modes as \cite{key-7}
\begin{equation}
\hat{\rho}_{boson}^{(R)}=\left(1-\exp\left(-8\pi M\omega\right)\right)\sum_{n}\exp\left(-8\pi nM\omega\right)|n_{out}^{(R)}><n_{out}^{(R)}|.\label{eq: matrice densita destra}
\end{equation}
Then, the average number of particles detected at infinity is \cite{key-7}

\begin{equation}
<n>_{boson}=tr\left[\hat{n}\hat{\rho}_{boson}^{(R)}\right]=\frac{1}{\exp\left(8\pi M\omega\right)-1},\label{eq: traccia}
\end{equation}
The trace (\ref{eq: traccia}) has been summed over all the eigenstates
and a bit of algebra has been used to obtain the final result, see
\cite{key-7} for details. Eq. (\ref{eq: traccia}) represents the
well known Bose-Einstein distribution. A similar analysis easily gives
the well known Fermi-Dirac distribution \cite{key-7}

\begin{equation}
<n>_{fermion}=\frac{1}{\exp\left(8\pi M\omega\right)+1},\label{eq: traccia 2}
\end{equation}
Both the distributions (\ref{eq: traccia}) and (\ref{eq: traccia 2})
represent a black body spectrum with the famous Hawking temperature
\cite{key-1,key-7}

\begin{equation}
T_{H}\equiv\frac{1}{8\pi M}.\label{eq: Hawking temperature}
\end{equation}
Thus, we shortly reviewed the result in \cite{key-7}, which is very
important. In fact, considering a reformulation of the tunnelling
mechanism, a perfect black body spectrum is found, in full agreement
with the original result by Hawking \cite{key-1}. On the other hand,
one immediately note that this result is in contrast with the result
in \cite{key-2,key-3}. The probability of emission which corresponds
with the two distributions (\ref{eq: traccia}) and (\ref{eq: traccia 2})
is indeed \cite{key-1,key-2,key-3} 
\begin{equation}
\Gamma\sim\exp(-\frac{\omega}{T_{H}}).\label{eq: hawking probability}
\end{equation}
But an exact calculation of the action for a tunnelling spherically
symmetric particle gives the important correction \cite{key-2,key-3}
\begin{equation}
\Gamma\sim\exp[-\frac{\omega}{T_{H}}(1-\frac{\omega}{2M})].\label{eq: Parikh Correction}
\end{equation}
This result is in contrast with the one in \cite{key-7} because it
releases the additional term $\frac{\omega}{2M}$ as a deviation from
the strict thermality \cite{key-2,key-3}. The key point is that in
\cite{key-7} the dynamical BH geometry due to the energy conservation
has not been taken into due account like in \cite{key-2,key-3}. The
energy conservation forces indeed the BH to contract during the emission
of the particle \cite{key-2,key-3}. In this way, the horizon recedes
from its original radius and becomes smaller at the end of the emission
process \cite{key-2,key-3}. Thus, BHs do not exactly emit like perfect
black bodies \cite{key-2,key-3}. 

In fact,  the tunnelling is a \emph{discrete} instead of \emph{continuous}
process \cite{key-8,key-10} because two different \emph{countable}
BH physical states have to be considered, the first before the emission
of the particle and the latter after the emission of the particle
\cite{key-8,key-10}. Then, the emission of the particle is interpreted
like a \emph{quantum} \emph{transition} of frequency $\omega$ between
the two different discrete states \cite{key-8,key-10}. The tunnelling
mechanism works indeed considering a trajectory in imaginary or complex
time joining two separated classical turning points \cite{key-2,key-3}.
The important consequence is that the radiation spectrum is also discrete
\cite{key-8,key-10}. Based on its importance, this issue has to be
clarified in a better way. If one fixes the Hawking temperature the
statistical probability distribution (\ref{eq: Parikh Correction})
is a continuous function. But the Hawking temperature in (\ref{eq: Parikh Correction})
varies in time with a behavior which is \emph{discrete.} The reason
of this discrete character is that the forbidden region traversed
by the emitting particle has a \emph{finite} size \cite{key-3}. Considering
a strictly thermal approximation, the turning points have zero separation.
Thus, it is not clear what joining trajectory one has to considered
because there is not barrier \cite{key-3}. The problem is solved
if one argues that it is the forbidden finite region from $r_{initial}=2M\:$
to $r_{final}=2(M\lyxmathsym{\textminus}\omega)\:$ that the tunnelling
particle traverses which acts like barrier \cite{key-3}. Then, the
intriguing explanation is that it is the particle itself which generates
a tunnel through the BH horizon \cite{key-3}.

If one wants to take into due account the dynamical geometry of the
BH during the emission of the particle a BH \emph{effective state
}can be introduced \cite{key-8}, \cite{key-10}-\cite{key-14}, \cite{key-20,key-46}.
In fact, introducing the \emph{effective temperature }\cite{key-8},
\cite{key-10}-\cite{key-14}, \cite{key-20,key-46} 
\begin{equation}
T_{E}(\omega)\equiv\frac{2M}{2M-\omega}T_{H}=\frac{1}{4\pi(2M-\omega)},\label{eq: Corda Temperature}
\end{equation}
eq. (\ref{eq: Corda Temperature}) can be rewritten in a Hawking-Boltzmann-like
form similar to the original probability found by Hawking 
\begin{equation}
\Gamma\sim\exp[-\beta_{E}(\omega)\omega]=\exp(-\frac{\omega}{T_{E}(\omega)}),\label{eq: Corda Probability}
\end{equation}

\noindent being $\exp[-\beta_{E}(\omega)\omega]$ the \emph{effective
Boltzmann factor }with \cite{key-8}, \cite{key-10}-\cite{key-14},
\cite{key-20,key-46}

\noindent 
\begin{equation}
\beta_{E}(\omega)\equiv\frac{1}{T_{E}(\omega)}.\label{eq: beta E}
\end{equation}
Therefore, the effective temperature replaces the Hawking temperature
in the equation of the probability of emission \cite{key-8}, \cite{key-10}-\cite{key-14},
\cite{key-20,key-46}. Let us discuss the physical interpretation
of $T_{E}(\omega)$ following \cite{key-20}. The probability of emission
of Hawking quanta found in \cite{key-2,key-3} i.e. eq. (\ref{eq: Parikh Correction}),
shows that the BH does NOT emit like a perfect black body, having
not a strictly thermal behavior. On the other hand, the temperature
in Bose-Einstein and Fermi-Dirac distributions is a perfect black
body temperature. Thus, when we have deviations from the strictly
thermal behavior, i.e. from the perfect black body, one expects also
deviations from Bose-Einstein and Fermi-Dirac distributions \cite{key-20}.
How can one attack this problem? By analogy with other various fields
of Science, also beyond BHs, for example the case of planets and stars
\cite{key-20}. One defines the effective temperature of a body such
as a star or planet as the temperature of a black body that would
emit the same total amount of electromagnetic radiation \cite{key-20,key-47}.
The importance of the effective temperature in a star is stressed
by the issue that the effective temperature and the bolometric luminosity
are the two fundamental physical parameters needed to place a star
on the Hertzsprung\textendash{}Russell diagram \cite{key-20}. Both
effective temperature and bolometric luminosity actually depend on
the chemical composition of a star, see again \cite{key-20,key-47}.\emph{
}The concept of effective temperature has been introduced by the author
in BH physics in \cite{key-10,key-11} for the Schwarzschild BH and
generalized in \cite{key-13} to the Kerr BH and in \cite{key-14}
to the non extremal Reissner-Nordström BH by the author and collaborators.
$T_{E}(\omega)$ depends on the energy-frequency of the emitted radiation
while the ratio $\frac{T_{E}(\omega)}{T_{H}}=\frac{2M}{2M-\omega}$
represents the deviation of the BH radiation spectrum from the strictly
thermal behavior \cite{key-8}, \cite{key-10}-\cite{key-14}, \cite{key-20,key-46}.

After the introduction of $T_{E}(\omega)$ one can introduce other
\emph{effective quantities.} Considering the initial BH mass \emph{before}
the emission, $M$, and the final BH mass \emph{after} the emission,
$M-\omega\:$ respectively \cite{key-8}, \cite{key-10}-\cite{key-14},
\cite{key-20,key-46}, the \emph{effective BH mass }and the \emph{effective
BH horizon} \emph{during} its contraction, i.e. \emph{during} the
emission of the particle are defined as \cite{key-8}, \cite{key-10}-\cite{key-14},
\cite{key-20,key-46} 
\begin{equation}
M_{E}\equiv M-\frac{\omega}{2},\mbox{ }r_{E}\equiv2M_{E}.\label{eq: effective quantities}
\end{equation}

\noindent These effective quantities are average quantities\emph{
}\cite{key-8}, \cite{key-10}-\cite{key-14}, \cite{key-20,key-46}.
The effective horizon \emph{$r_{E}$ }is the average of the initial
and final horizons and the effective mass \emph{$M_{E}\:$ }is the
average of the initial and final masses \cite{key-8}, \cite{key-10}-\cite{key-14},
\cite{key-20,key-46}. \emph{Before} the emission the Hawking temperature
is $T_{H\mbox{ initial}}=\frac{1}{8\pi M}$ ; \emph{after} the emission
one gets $T_{H\mbox{ final}}=\frac{1}{8\pi(M-\omega)}.$ Then, \emph{$T_{E}\:$
}results to be the inverse of the average value of the inverses of
the initial and final Hawking temperatures \cite{key-8}, \cite{key-10}-\cite{key-14},
\cite{key-20,key-46}. This implies that the Hawking temperature \emph{has
a discrete character in time} \cite{key-8}, \cite{key-10}-\cite{key-14}.
We stress that the introduction of the effective temperature does
not degrade the importance of the Hawking temperature \cite{key-20}.
In fact, as $T_{H}$ changes with a discrete behavior in time, let
us ask: which value of the Hawking temperature has to be associated
to the emission of the particle? Has one to consider the value of
the Hawking temperature \emph{before} the \emph{emission} or the value
of the Hawking temperature after the emission? The answer is that
one has to consider an \emph{intermediate} value, the effective temperature,
which is the inverse of the average value of the inverses of the initial
and final Hawking temperatures \cite{key-20}. In some way, $\: T_{E}(\omega)$
represents the value of the Hawking temperature \emph{during} the
emission of the quantum \cite{key-8}, \cite{key-10}-\cite{key-14},
\cite{key-20,key-46}. Then, the effective temperature takes into
account the non-strictly thermal character of the radiation spectrum
and the non-strictly continuous character of subsequent emissions
of Hawking quanta.

Now, let us rewrite eq. (\ref{eq: beta E}) as \cite{key-8} 
\begin{equation}
\beta_{E}(\omega)\equiv\frac{1}{T_{E}(\omega)}=\beta_{H}\left(1-\frac{\omega}{2M}\right),\label{eq: beta E-1}
\end{equation}
where $\beta_{H}\equiv\frac{1}{T_{H}}$. Using Hawking\textquoteright{}s
arguments \cite{key-8,key-23,key-24} , we write down the euclidean
form of the metric as \cite{key-8}

\begin{equation}
ds_{E}^{2}=x^{2}\left[\frac{d\tau}{4M\left(1-\frac{\omega}{2M}\right)}\right]^{2}+\left(\frac{r}{r_{E}}\right)^{2}dx^{2}+r^{2}(\sin^{2}\theta d\varphi^{2}+d\theta^{2}),\label{eq: euclidean form}
\end{equation}
which is regular at $x=0$ and $r=r_{E}$. $\tau$ is treated as an
angular variable with period $\beta_{E}(\omega)$ \cite{key-8,key-23,key-24}.
Replacing the quantity $\sum_{i}\beta_{i}\frac{\hslash^{i}}{M^{2i}}$
in \cite{key-23} with the quantity $-\frac{\omega}{2M}$ \cite{key-8}
one follows step by step the detailed analysis in \cite{key-23} obtaining
the \emph{effective Schwarzschild line element} \cite{key-8} 
\begin{equation}
ds_{E}^{2}\equiv-(1-\frac{2M_{E}}{r})dt^{2}+\frac{dr^{2}}{1-\frac{2M_{E}}{r}}+r^{2}(\sin^{2}\theta d\varphi^{2}+d\theta^{2}).\label{eq: Hilbert effective}
\end{equation}
It is also simple to check that $r_{E}$ in eq. (\ref{eq: euclidean form})
is the same as in eq. (\ref{eq: effective quantities}) \cite{key-8}.
Therefore, the \emph{effective} \emph{surface gravity} can be defined
as\emph{ }$\frac{1}{4M_{E}}.$ Thus, the effective line element (\ref{eq: Hilbert effective})
takes into account the BH\emph{ dynamical} geometry during the emission
of the particle. Following step by step the analysis in \cite{key-7},
at the end the correct physical states for boson and fermions read
\cite{key-8} 
\begin{equation}
\begin{array}{c}
|\Psi>_{boson}=\left(1-\exp\left(-8\pi M_{E}\omega\right)\right)^{\frac{1}{2}}\sum_{n}\exp\left(-4\pi nM_{E}\omega\right)|n_{out}^{(L)}>\otimes|n_{out}^{(R)}>\\
\\
|\Psi>_{fermion}=\left(1+\exp\left(-8\pi M_{E}\omega\right)\right)^{-\frac{1}{2}}\sum_{n}\exp\left(-4\pi nM_{E}\omega\right)|n_{out}^{(L)}>\otimes|n_{out}^{(R)}>.
\end{array}\label{eq: physical states-1}
\end{equation}
Then, one immediately finds that the correct, non-strictly thermal,
distributions are given exactly by eq. (\ref{eq: final distributions}).
Those equations represent the distributions associated to probability
of emission (\ref{eq: Parikh Correction}).

Resuming, in \cite{key-7} the tunnelling approach on Hawking radiation
has been improved by explicitly finding a black body spectrum associated
with BHs. This result has been obtained by using a reformulation of
the tunnelling mechanism. On the other hand, the result in \cite{key-7}
had a problem because it was in contrast which the result in \cite{key-2,key-3},
that found a probability of emission compatible with a non-strictly
black body spectrum instead. Using the recent introduction of a BH
effective state \cite{key-8}, \cite{key-10}-\cite{key-14}, \cite{key-20,key-46}
such a contradiction has been solved in \cite{key-8} through a slight
modification of the analysis in \cite{key-7}. In this Section the
analysis in \cite{key-8} has been reviewed, showing that the final
result consists in a non-strictly black body spectrum from the tunnelling
mechanism with the two associated non-strictly distributions (\ref{eq: final distributions}).

\section{Quasi-normal modes as ``electron states'' in a Bohr-like black
hole model }

\noindent In this Section we review the results in \cite{key-10,key-11,key-12}
for the Schwarzschild BH showing that the correction to the thermal
spectrum in \cite{key-2,key-3} is also very important for the physical
interpretation of BH QNMs and, in turn, it results important to realize
the underlying theory of quantum gravity. It is indeed a general conviction
that BHs are theoretical laboratories for developing a quantum gravity
theory and in this paper BH QNMs are naturally interpreted in terms
of quantum levels, the ``electron states'' of a Bohr-like BH model
\cite{key-10,key-11,key-12}. 

BH QNMs are modes of radial perturbations obeying the time independent
Schröedinger-like equation \cite{key-10,key-11,key-12,key-26} 
\begin{equation}
\left(-\frac{\partial^{2}}{\partial x^{2}}+V(x)-\omega^{2}\right)\phi.\label{eq: diff.}
\end{equation}
Working in strictly thermal approximation the Regge-Wheeler potential
is introduced as \cite{key-10,key-11,key-12,key-26}

\noindent 
\begin{equation}
V(x)=V\left[x(r)\right]=\left(1-\frac{2M}{r}\right)\left(\frac{l(l+1)}{r^{2}}+2\frac{(1-j^{2})M}{r^{3}}\right),\label{eq: Regge-Wheeler-1}
\end{equation}
where $j=0,1,2$ for scalar, vector and gravitational perturbation
respectively.

\noindent The relation between the Regge-Wheeler ``tortoise'' coordinate
$x$ and the radial coordinate $r\;$ is \cite{key-10,key-11,key-12,key-26}

\noindent 
\begin{equation}
\begin{array}{c}
x=r+2M\ln\left(\frac{r}{2M}-1\right)\\
\\
\frac{\partial}{\partial x}=\left(1-\frac{2M}{r}\right)\frac{\partial}{\partial r}.
\end{array}\label{eq:original  tortoise}
\end{equation}

\noindent BH quasi-normal frequencies are analogous to quasi-stationary
states in quantum mechanics \cite{key-12,key-26} . In that way, their
frequency can be complex \cite{key-12,key-26}. Purely outgoing boundary
conditions are requested both at the horizon ($r=2M$) and in the
asymptotic region ($r\rightarrow\infty$) \cite{key-12,key-26}

\noindent 
\begin{equation}
\phi(x)\sim c_{\pm}\exp\left(\mp i\omega x\right)\quad for\quad x=\pm\infty.\label{eq: condizioni contorno}
\end{equation}
The intriguing idea to model the quantum BH in terms of QNMs was pioneered
by York in \cite{key-27}. In that work, using the statistical mechanics
of the QNMs, the value of $\left((0.27654)\right)\left(16\pi M^{2}\right)$
was obtained for the BH entropy, which is near the value usually assumed
of the standard Bekenstein-Hawking entropy $4\pi M^{2}$. \emph{Considering
Bohr's Correspondence Principle }\cite{key-45},\emph{ }which states
that \textquotedblleft{}\emph{transition frequencies at large quantum
numbers should equal classical oscillation frequencies}\textquotedblright{},
in \cite{key-28,key-29} the important result that QNMs release information
about the area quantization was obtained by Hod. The idea in \cite{key-28,key-29}
was to consider the QNMs as associated to absorption of particles.
Some important problems arising from this approach were solved in
\cite{key-25} by Maggiore. An important issue is that QNMs are \emph{countable}
frequencies. This was in contrast with ideas on the \emph{continuous}
character of Hawking radiation, preventing attempts to interpret QNMs
in terms of emitted quanta, and, in turn, to associate QNMs modes
to Hawking radiation \cite{key-26}. On the other hand, the authors
of \cite{key-15,key-16,key-17,key-18,key-19} and ourselves and collaborators
\cite{key-10}-\cite{key-14}, \cite{key-20,key-46} made the important
observation that the non-thermal spectrum in \cite{key-2,key-3} also
implies the countable character of subsequent emissions of Hawking
quanta. This key point enables a natural correspondence between Hawking
radiation and BH QNMs \cite{key-10}-\cite{key-14}, \cite{key-20,key-46}
and permits, in turn, to interpret QNMs also in terms of emitted energies\cite{key-10}-\cite{key-14},
\cite{key-20,key-46}. BH QNMs represent indeed the BH reaction to
small, discrete perturbations in terms of damped oscillations \cite{key-10}-\cite{key-14},
\cite{key-20,key-46}. If the capture of a particle causing an increase
in the horizon area is a type of discrete perturbation \cite{key-25,key-28,key-29},
it is obvious and natural to consider the emission of a particle causing
a decrease in the horizon area also as a perturbation which generates
a reaction in terms of countable QNMs, being that process discrete
rather than continuous \cite{key-10}-\cite{key-14}, \cite{key-20,key-46}.
This can be immediately understood if one considers that it is the
particle having negative energy and tunnelling inwards which is captured
by the hole in the mechanism of pair creation. Thus, the absorbed
Hawking quanta having negative energies generate subsequent perturbations
``triggering'' the QNMs. Concerning this key point, it is important
to emphasize that the correspondence existing between emitted radiation
and proper oscillation of the emitting body is well known as a fundamental
behavior of every radiation process in Science. This natural correspondence
between Hawking radiation and BH QNMs, permits immediately and naturally
to interpret BH QNMs in terms of quantum levels also for emitted energies
\cite{key-10}-\cite{key-14}, \cite{key-20,key-46}. This is also
in agreement with the general idea that BHs can be considered in terms
of highly excited states representing both the ``hydrogen atom''
and the ``quasi-thermal emission'' in an underlying theory of quantum
gravity \cite{key-12,key-28,key-29}. 

\noindent Considering a strictly thermal approximation, BH QNMs are
usually labelled as $\omega_{nl},$ being $l$ the angular momentum
quantum number \cite{key-10,key-11,key-12,key-25,key-26,key-28,key-29}.
For each $l$, one finds a countable sequence of BH QNMs, labelled
by the principal quantum number $n$ ($n=1,2,...$) \cite{key-10,key-11,key-12,key-25,key-26,key-28,key-29}.
For large $n$ the Schwarzschild BH QNMs result independent of $l$.
They have the structure \cite{key-4,key-5,key-10,key-11,key-12}

\noindent 
\begin{equation}
\begin{array}{c}
\omega_{n}=\ln3\times T_{H}+2\pi i(n+\frac{1}{2})\times T_{H}+\mathcal{O}(n^{-\frac{1}{2}})=\\
\\
=\frac{\ln3}{8\pi M}+\frac{2\pi i}{8\pi M}(n+\frac{1}{2})+\mathcal{O}(n^{-\frac{1}{2}}).
\end{array}\label{eq: quasinormal modes}
\end{equation}

\noindent Eq. (\ref{eq: quasinormal modes}) has been originally obtained
numerically in \cite{key-30,key-31}. Later, it has been also derived
in analytical way in \cite{key-26,key-32}. On the other hand, eq.
(\ref{eq: quasinormal modes}) is strictly correct only for scalar
and gravitational perturbations \cite{key-12,key-25,key-26}. In any
case, as we work in the large $n$ approximation, the leading term
in the imaginary part of the complex frequencies becomes dominant
\cite{key-12,key-25,key-26}, in full agreement with Bohr correspondence
principle \cite{key-45}. Then, eq. (\ref{eq: quasinormal modes})
is well approximated by \cite{key-12,key-25,key-26}

\begin{equation}
\omega_{n}\simeq\frac{2\pi in}{8\pi M}\label{eq: well approximated}
\end{equation}
In \cite{key-26} it has been shown that eq. (\ref{eq: well approximated})
holds also for vector and half integer spin perturbations, again in
agreement with Bohr correspondence principle. A key point is that
eq. (\ref{eq: well approximated}) works only in strictly thermal
approximation. If one wants to take into due account the deviation
from the perfect black body spectrum the Hawking temperature $T_{H}$
must be replaced by the effective temperature $T_{E}$ in eq. (\ref{eq: well approximated})
obtaining \cite{key-10,key-11,key-12} 
\begin{equation}
\omega_{n}\simeq\frac{2\pi in}{4\pi\left[2M-|\omega_{n}|\right]}.\label{eq: andamento asintotico}
\end{equation}
An intuitive derivation of eq. (\ref{eq: andamento asintotico}) can
be found in \cite{key-10,key-11}. We rigorously derived such an equation
in the Appendix of \cite{key-12}. Further details on that derivation
can be analysed as it follows. The BH\emph{ dynamical} geometry during
the emission of the particle is taken into account by the effective
line element (\ref{eq: Hilbert effective}). Although this does not
mean that one can immediately replace $T_{H}(M)$ with $T_{H}(M-\frac{\omega}{2})$
in eq. (\ref{eq: well approximated}), the effective line element
(\ref{eq: Hilbert effective}) permits to introduce the \emph{effective
equations} \cite{key-10,key-11,key-12} 
\begin{equation}
\left(-\frac{\partial^{2}}{\partial x^{2}}+V(x)-\omega^{2}\right)\phi,\label{eq: diff.-1}
\end{equation}
 
\begin{equation}
V(x)=V\left[x(r)\right]=\left(1-\frac{2M_{E}}{r}\right)\left(\frac{l(l+1)}{r^{2}}+2\frac{(1-j^{2})M_{E}}{r^{3}}\right),\label{eq: effettiva 1}
\end{equation}
where $V\left[x(r)\right]$ is the \emph{effective Regge-Wheeler potential},
and 
\begin{equation}
\begin{array}{c}
x=r+2M_{E}\ln\left(\frac{r}{2M_{E}}-1\right)\\
\\
\frac{\partial}{\partial x}=\left(1-\frac{2M_{E}}{r}\right)\frac{\partial}{\partial r}.
\end{array}\label{eq: effettiva 2}
\end{equation}
In order to simplify the following equations, here we also set \cite{key-12}
\begin{equation}
2M_{E}=r_{E}\equiv1\;\;\;\; and\;\;\;\; m\equiv n+1.\label{eq: set}
\end{equation}

\noindent We stress that the \emph{Planck mass} $m_{p}\:$ is equal
to $1$ in Planck units. Then, one rewrites (\ref{eq: quasinormal modes})
as \cite{key-12} 
\begin{equation}
\frac{\omega_{m}}{m_{p}^{2}}=\frac{\ln3}{4\pi}+\frac{i}{2}(m-\frac{1}{2})+\mathcal{O}(m^{-\frac{1}{2}}),\quad for\: m\gg1,\label{eq: intuizione}
\end{equation}
where now $m_{p}\neq1.$ Setting \cite{key-12}

\noindent 
\begin{equation}
\tilde{\omega}_{m}\equiv\frac{\omega_{m}}{m_{p}^{2}},\label{eq: important definition}
\end{equation}

\noindent eqs. (\ref{eq: intuizione}), (\ref{eq: diff.-1}), (\ref{eq: effettiva 1})
and (\ref{eq: effettiva 2}) read \cite{key-12} 
\begin{equation}
\tilde{\omega}_{m}=\frac{\ln3}{4\pi}+\frac{i}{2}(m-\frac{1}{2})+\mathcal{O}(m^{-\frac{1}{2}}),\quad for\: m\gg1,\label{eq: rottura}
\end{equation}

\noindent 
\begin{equation}
\left(-\frac{\partial^{2}}{\partial x^{2}}+V(x)-\tilde{\omega}^{2}\right)\phi,\label{eq: rottura 0}
\end{equation}

\noindent 
\begin{equation}
V(x)=V\left[x(r)\right]=\left(1-\frac{1}{r}\right)\left(\frac{l(l+1)}{r^{2}}-\frac{3(1-j^{2})}{r^{3}}\right)\label{eq: rottura 2}
\end{equation}

\noindent and 
\begin{equation}
\begin{array}{c}
x=r+\ln\left(r-1\right)\\
\\
\frac{\partial}{\partial x}=\left(1-\frac{1}{r}\right)\frac{\partial}{\partial r}
\end{array}\label{eq: rottura 3}
\end{equation}
respectively. Now, if one replies the same rigorous analytical calculation
in the Appendix of \cite{key-12} or the analogous calculation in
\cite{key-26}, but starting from eqs. (\ref{eq: rottura 0}), (\ref{eq: rottura 2})
and (\ref{eq: rottura 3}) and satisfying purely outgoing boundary
conditions both at the effective horizon ($r_{E}=2M_{E}$) and in
the asymptotic region ($r=\infty$), the final result in the leading
term in the imaginary part of the complex frequencies will be, obviously
and rigorously, eq. (\ref{eq: andamento asintotico}). An important
issue has to be clarified. One could take position against the above
analysis claiming that $M_{E}$ and $r_{E}$ (and consequently the
\emph{effective tortoise coordinate} and the effective Regge-Wheeler
potential) are frequency dependent. But we note that eq. (\ref{eq: set})
translates such a frequency dependence into a continually rescaled
mass unit in the discussion in the Appendix of \cite{key-12}. It
is simple to show that such a rescaling is extremely slow and always
included within a factor $2.$ Thus, it does not influence the analysis
in the Appendix of \cite{key-12}. In fact, we note that, although
$\tilde{\omega}$ in the analysis in the Appendix of \cite{key-12}
can be very large because of definition (\ref{eq: important definition}),
$\omega$ must instead be always minor than the BH initial mass as
BHs cannot emit more energy than their total mass. Thus, the analysis
in the Appendix of \cite{key-12} can be considered a ``quasi-asymptotic''
analysis, i.e the $\omega$ can be extremely large but not infinity,
see also the below discussion on the maximum value for the overtone
number $n.$ Inserting this constraint in eq. (\ref{eq: effective quantities})
one gets the range of permitted values of $M_{E}(|\omega_{n}|)$ as
\cite{key-12}

\begin{equation}
\frac{M}{2}\leq M_{E}(|\omega_{n}|)\leq M.\label{eq: range 1}
\end{equation}
Thus, setting $2M_{E}(|\omega_{n}|)=r_{E}(|\omega_{n}|)\equiv1(|\omega_{n}|)$
one sees that the range of permitted values of the continually rescaled
mass unit is always included within a factor $2$ \cite{key-12}.
On the other hand, the countable sequence of QNMs is very large, see
the below discussion on the maximum value for the overtone number
$n$ and \cite{key-10,key-13}. This implies that the mass unit's
rescaling is extremely slow \cite{key-12}. Therefore, the reader
can easily check, by reviewing the discussion in the Appendix of \cite{key-12}
step by step, that the continually rescaled mass unit did not influence
the analysis.

Let us discuss another argument which emphasizes the correctness of
the analysis in the Appendix of \cite{key-12}. We can choose to consider
$M_{E}$ as being constant within the range (\ref{eq: range 1}) \cite{key-12}.
In that case, we show that such an approximation is very good \cite{key-12}.
Eq. (\ref{eq: range 1}) implies indeed that the range of permitted
values of $T_{E}(|\omega_{n}|)$ is \cite{key-12} 
\begin{equation}
T_{H}=T_{E}(0)\leq T_{E}(|\omega_{n}|)\leq2T_{H}=T_{E}(|\omega_{n_{max}}|),\label{eq: range 2}
\end{equation}
where $T_{H}$ is the initial Hawking temperature. Then, if we fix
$M_{E}=\frac{M}{2}$ in the analysis, the approximate result is \cite{key-12}

\begin{equation}
\omega_{n}\simeq2\pi in\times2T_{H}.\label{eq: approssima 1}
\end{equation}
On the other hand, if one fixes $M_{E}=M\:$ as in thermal approximation,
the approximate result is \cite{key-12}

\begin{equation}
\omega_{n}\simeq2\pi in\times T_{H}.\label{eq: approssima 2}
\end{equation}
We see that both the approximate results in correspondence of the
extreme values in the range (\ref{eq: range 1}) have the same order
of magnitude \cite{key-12}. Thus, fixing $2M_{E}=r_{E}\equiv1$ does
not change the order of magnitude of the final (approximated) result
with respect to the exact result \cite{key-12}. In particular, setting
$T_{E}=\frac{3}{2}T_{H}$ the uncertainty in the final result is $0.33$,
while in the result of the thermal approximation (\ref{eq: approssima 2})
the uncertainty is $2$ \cite{key-12}. Hence, even if one considers
$M_{E}$ as constant, the result in the Appendix of \cite{key-12}
is more precise than the thermal approximation of previous literature.
Thus, the derivation of eq. (\ref{eq: andamento asintotico}) is surely
correct. 

\noindent Eq. (\ref{eq: andamento asintotico}) has the following
elegant interpretation \cite{key-10,key-11}. QNMs determine the position
of poles of a Green's function on the given background and the Euclidean
BH solution converges to a \emph{non-strictly} thermal circle at infinity
with the inverse temperature $\beta_{E}(\omega_{n})=\frac{1}{T_{E}(|\omega_{n}|)}$
\cite{key-10,key-11}. Then, the spacing of the poles in eq. (\ref{eq: andamento asintotico})
coincides with the spacing $2\pi iT_{E}(|\omega_{n}|)=2\pi iT_{H}(\frac{2M}{2M-|\omega_{n}|}),$
expected for a \emph{non-strictly} thermal Green's function \cite{key-10,key-11}. 

\noindent As BHs cannot emit more energy than their total mass, the
physical solution for the absolute values of the frequencies (\ref{eq: andamento asintotico}),
i.e. the one for which it is $|\omega_{n}|\leq M$, is \cite{key-10,key-11,key-12}

\noindent 
\begin{equation}
E_{n}\equiv|\omega_{n}|=M-\sqrt{M^{2}-\frac{n}{2}}.\label{eq: radice fisica}
\end{equation}
$E_{n}$ is interpreted like the total energy emitted at level $n$
\cite{key-10,key-11,key-12}. As the square root in eq. (\ref{eq: radice fisica})
must be a real non negative number, we need also \cite{key-10,key-12}

\begin{equation}
M^{2}-\frac{n}{2}\geq0.\label{eq: need}
\end{equation}
Solving the expression (\ref{eq: need}) one gets a maximum value
for the overtone number $n$ 

\begin{equation}
n\leq n_{max}=2M^{2},\label{eq: n max}
\end{equation}
corresponding to $E_{n}=M.$ This means that $n_{max}$ is a finite
number, although it can be very very large. The result (\ref{eq: n max})
is correct if one assumes a total BH evaporation. But in \cite{key-33}
it has been shown that the Generalized Uncertainty Principle prevents
the total BH evaporation in exactly the same way that the Uncertainty
Principle prevents the hydrogen atom from total collapse. In fact,
the collapse is prevented by dynamics rather than by symmetry, when
the \emph{Planck distance} and the \emph{Planck mass} are approached
\cite{key-33}. Then, one has to slightly modify eq. (\ref{eq: need})
obtaining (the \emph{Planck mass} is equal to $1$ in Planck units) 

\begin{equation}
M^{2}-\frac{n}{2}\geq1.\label{eq: need 1}
\end{equation}
The solution of eq. (\ref{eq: need 1}) is 

\begin{equation}
n\leq n_{max}=2(M^{2}-1).\label{eq: n max 1}
\end{equation}
which gives a different value of the maximum value for the overtone
number $n.$ Let us consider, for example, a BH mass of 10 Solar masses.
One obtains $n_{max}\sim10^{78}$ from both eqs. (\ref{eq: n max})
and (\ref{eq: n max 1}). Thus, we understand that, although the total
number of QNMs for emitted energies is not infinity, our ``quasi-asymptotic''
analysis for large $n$ is an excellent approximation.

\noindent Considering an emission from the ground state (i.e. a BH
which is not excited) to a state with large $n=n_{1}$ and using eq.
(\ref{eq: radice fisica}), the BH mass changes from $M\:$ to 

\begin{equation}
M_{n_{1}}\equiv M-E_{n_{1}}=\sqrt{M^{2}-\frac{n_{1}}{2}}.\label{eq: me n1}
\end{equation}
In the transition from the state with $n=n_{1}$ to a state with $n=n_{2}$
where $n_{2}>n_{1}$ the BH mass changes again from $M_{n_{1}}\:$
to

\begin{equation}
\begin{array}{c}
M_{n_{2}}\equiv M_{n_{1}}-\Delta E_{n_{1}\rightarrow n_{2}}=M-E_{n_{2}}\\
=\sqrt{M^{2}-\frac{n_{2}}{2}},
\end{array}\label{eq: me n2}
\end{equation}
where 
\begin{equation}
\Delta E_{n_{1}\rightarrow n_{2}}\equiv E_{n_{2}}-E_{n_{1}}=M_{n_{1}}-M_{n_{2}}=\sqrt{M^{2}-\frac{n_{1}}{2}}-\sqrt{M^{2}-\frac{n_{2}}{2}},\label{eq: jump}
\end{equation}
is the jump between the two levels due to the emission of a particle
having frequency $\Delta E_{n_{1}\rightarrow n_{2}}$. Thus, we have
found the intriguing result that the BH mass varies in function of
the initial mass $M$ and of the BH quantum level \cite{key-12}. 

Now, following \cite{key-10,key-11,key-12}, important consequences
on BHs quantum physics, which arise from the correspondence between
Hawking radiation and BH QNMs, will be discussed, starting from the
\emph{area quantization}. 

\noindent Bekenstein \cite{key-34} proposed that the area of the
BH horizon is quantized in units of the Planck length in quantum gravity
(the \emph{Planck length} $l_{p}=1.616\times10^{-33}\mbox{ }cm$ is
equal to one in Planck units). His famous result was that the Schwarzschild
BH area quantum is $\triangle A=8\pi$ \cite{key-34}. In \cite{key-28,key-29}
Hod had the intriguing idea to consider BH QNMs like quantum levels
for absorption of particles. In that way, he found a different numerical
coefficient \cite{key-28,key-29}. It is important to recall \cite{key-54}
that the Hod rule $\Delta A=4\ln3$ is actually a special case of
one suggested by Mukhanov in \cite{key-55}. Mukhanov proposed $\Delta A=4\ln k\quad with\quad k=2,3,...$.
This entered into the joint paper of Bekenstein and Mukhanov \cite{key-56}
and then into the review of Bekenstein \cite{key-53}. In any case,
Hod's work was reanalyzed in \cite{key-25} by Maggiore, who re-obtained
the original result of Bekenstein. We further improved the result
in \cite{key-25} taking into account the deviation from the perfect
thermal spectrum in \cite{key-10,key-11,key-12} and adding to the
analysis the perturbations due to the subsequent emissions of Hawking
quanta. In fact, in our approach we used eq. (\ref{eq: andamento asintotico})
instead of eq. (\ref{eq: well approximated}) \cite{key-10,key-11,key-12}.
Setting $n_{1}=n-1$, $n_{2}=n\:$ in eq. (\ref{eq: jump}), one gets
the emitted energy for a jump among two neighboring levels \cite{key-10,key-11,key-12}
\begin{equation}
\Delta E_{n-1\rightarrow n}=E_{n}-E_{n-1}=\sqrt{M^{2}-\frac{n-1}{2}}-\sqrt{M^{2}-\frac{n}{2}}\label{eq: variazione}
\end{equation}
The result (\ref{eq: variazione}) holds for Kerr BHs too, when $M^{2}\gg J,$
where $J$ is the angular momentum of the BH \cite{key-13}. In the
Schwarzschild BH the \emph{horizon area} $A$ is related to the mass
by the relation $A=16\pi M^{2}.$ Thus, a variation $\triangle M\,$
of the mass implies a variation

\noindent 
\begin{equation}
\triangle A=32\pi M\triangle M\label{eq: variazione area}
\end{equation}
of the area. On the other hand, after an high number of emissions
(and potential absorptions because neighboring particles can be can
captured by the BH), the BH mass changes from $M\;$ to \cite{key-12}

\begin{equation}
M_{n-1}\equiv M-E_{n-1},\label{eq: me-1}
\end{equation}
where $E_{n-1}$ is the total energy emitted by the BH at that time
(the BH is excited at a level $n-1$ \cite{key-12}). A further emission
causes a transition from the state with $n-1$ to the state with $n\;$
\cite{key-12}, and the BH mass changes again from $M_{n-1}$ to \cite{key-12}

\begin{equation}
M_{n}\equiv M-E_{n-1}-\Delta E_{n-1\rightarrow n}.\label{eq: me}
\end{equation}
If one uses eq. (\ref{eq: variazione}) one gets \cite{key-12} 
\begin{equation}
\begin{array}{c}
M_{n}=M-E_{n-1}+\sqrt{M^{2}-\frac{n}{2}}-\sqrt{M^{2}-\frac{n-1}{2}}\\
\\
=M-E_{n-1}+E_{n-1}-E_{n}=M-E_{n},
\end{array}\label{eq: me 2}
\end{equation}
and now the BH is excited at the level $n$ \cite{key-12}. Using
eq. (\ref{eq: radice fisica}), eqs. (\ref{eq: me-1}) and (\ref{eq: me 2})
become \cite{key-12} 
\begin{equation}
M_{n-1}=\sqrt{M^{2}-\frac{n-1}{2}},\label{eq: me 3}
\end{equation}
and 

\begin{equation}
M_{n}=\sqrt{M^{2}-\frac{n}{2}}.\label{eq: me 4}
\end{equation}
Then, using eqs. (\ref{eq: variazione area}) and eq. (\ref{eq: variazione})
we get \cite{key-12}

\begin{equation}
\triangle A_{n-1}\equiv32\pi M_{n-1}\Delta E_{n-1\rightarrow n}\label{eq: area quantum e}
\end{equation}
Eq. (\ref{eq: area quantum e}) should give the area quantum of an
excited BH when one considers an emission from the level $n-1$ to
the level $n$ in function of the principal quantum number $n$ and
of the initial BH mass. Actually, there is a problem. In fact, an
absorption from the level $n$ to the level $n-1$ is now possible
and the correspondent absorbed energy is \cite{key-11,key-12} 
\begin{equation}
E_{n-1}-E_{n}=-\Delta E_{n-1\rightarrow n}\equiv\Delta E_{n\rightarrow n-1}.\label{eq: absorbed}
\end{equation}
Then, the area quantum for the transition (\ref{eq: absorbed}) should
be 
\begin{equation}
\triangle A_{n}\equiv32\pi M_{n}\Delta E_{n\rightarrow n-1}\label{eq: area quantum a}
\end{equation}
and one gets the strange result that the absolute value of the area
quantum for an emission from the level $n-1$ to the level $n\;$
is different from the absolute value of the area quantum for an absorption
from the level $n$ to the level $n-1$ because it is $M_{n-1}\neq M_{n}$
\cite{key-12}. One expects the area spectrum to be the same for absorption
and emission instead \cite{key-12}. In order to solve this inconsistency
one considers again the \emph{effective mass} corresponding to the
transitions between the two levels $n\;$ and $n-1$. In fact, that
effective mass is the same for emission and absorption \cite{key-12}

\begin{equation}
\begin{array}{c}
M_{E(n,\; n-1)}\equiv\frac{1}{2}\left(M_{n-1}+M_{n}\right)\\
\\
\frac{1}{2}\left(\sqrt{M^{2}-\frac{n-1}{2}}+\sqrt{M^{2}-\frac{n}{2}}\right).
\end{array}\label{eq: massa effettiva n}
\end{equation}
If one replaces $M_{n-1}$ with $M_{E(n,\; n-1)}$ in eq. (\ref{eq: area quantum e})
and $M_{n}$ with $M_{E(n,\; n-1)}$ in eq. (\ref{eq: area quantum a})
one obtains 
\begin{equation}
\begin{array}{c}
\triangle A_{n-1}\equiv32\pi M_{E(n,\; n-1)}\,\Delta E_{n-1\rightarrow n}\qquad emission\\
\\
\triangle A_{n}\equiv32\pi M_{E(n,\; n-1)}\,\Delta E_{n\rightarrow n-1}\qquad absorption
\end{array}\label{eq: expects}
\end{equation}
and now it is $|\triangle A_{n}|=|\triangle A_{n-1}|.$ By using eqs.
(\ref{eq: variazione}) and (\ref{eq: massa effettiva n}) one finds 

\begin{equation}
|\triangle A_{n}|=|\triangle A_{n-1}|=8\pi,\label{eq: 8 pi planck}
\end{equation}
which is exactly the original famous result by Bekenstein \cite{key-34},
which is spin independent and in full agreement with Bohr's correspondence
principle \cite{key-12}. Thus, one takes the result (\ref{eq: 8 pi planck})
as the quantization of the area of the horizon of a Schwarzschild
BH. This is a strong confirmation of the correctness of our analysis.
Putting $A_{n-1}\equiv16\pi M_{n-1}^{2}$ and $A_{n}\equiv16\pi M_{n}^{2}$,
the formulas of the number of quanta of area are \cite{key-12}

\noindent 
\begin{equation}
N_{n-1}\equiv\frac{A_{n-1}}{|\triangle A_{n-1}|}=\frac{16\pi M_{n-1}^{2}}{32\pi M_{E(n,\; n-1)}\cdot\Delta E_{n-1\rightarrow n}}=\frac{M_{n-1}^{2}}{2M_{E(n,\; n-1)}\cdot\Delta E_{n-1\rightarrow n}}\label{eq: N n-1}
\end{equation}

\noindent before the emission, and \cite{key-12}
\begin{equation}
N_{n}\equiv\frac{A_{n}}{|\triangle A_{n}|}=\frac{16\pi M_{n}^{2}}{32\pi M_{E(n,\; n-1)}\cdot\triangle M_{n}}=\frac{M_{n}^{2}}{2M_{E(n,\; n-1)}\cdot\Delta E_{n-1\rightarrow n}}\label{eq: N n}
\end{equation}
after the emission respectively. It is easily to check that \cite{key-12}
\begin{equation}
N_{n-1}-N_{n}=\frac{M_{n-1}^{2}-M_{n}^{2}}{2M_{E(n,\; n-1)}\cdot\Delta E_{n-1\rightarrow n}}=\frac{\triangle M_{n}\left(M_{n-1}+M_{n}\right)}{2M_{E(n,\; n-1)}\cdot\Delta E_{n-1\rightarrow n}}=1\label{eq: check}
\end{equation}
as one expects. Then, the formulas of the famous Bekenstein-Hawking
entropy are \cite{key-12} 
\begin{equation}
\left(S_{BH}\right)_{n-1}\equiv\frac{A_{n-1}}{4}=8\pi N_{n-1}M_{n-1}\cdot\Delta E_{n-1\rightarrow n}=4\pi\left(M^{2}-\frac{n-1}{2}\right)\label{eq: Bekenstein-Hawking  n-1}
\end{equation}
before the emission and \cite{key-12} 
\begin{equation}
\left(S_{BH}\right)_{n}\equiv\frac{A_{n}}{4}=8\pi N_{n}M_{n}\cdot\Delta E_{n-1\rightarrow n}=4\pi\left(M^{2}-\frac{n}{2}\right)\label{eq: Bekenstein-Hawking  n}
\end{equation}
after the emission respectively. Notice that, as $n\gg1$, one obtains
$\left(S_{BH}\right)_{n}\simeq\left(S_{BH}\right)_{n-1}$ \cite{key-12}.
Formulas (\ref{eq: Bekenstein-Hawking  n-1}) and (\ref{eq: Bekenstein-Hawking  n})
are extremely important. In fact, they give the Bekenstein-Hawking
entropy as function of the BH original mass and of the BH quantum
level $n.$ They work for all $j=0,1,2,$ in total agreement with
Bohr's correspondence principle. 

On the other hand, the total BH entropy contains at least three parts
which are necessary to realize the underlying theory of quantum gravity
\cite{key-10,key-11,key-12,key-35,key-36}. They are the usual Bekenstein-Hawking
entropy and two sub-leading corrections: the logarithmic term and
the inverse area term \cite{key-10,key-11,key-12,key-35,key-36}

\noindent 
\begin{equation}
S_{total}=S_{BH}-\ln S_{BH}+\frac{3}{2A}.\label{eq: entropia totale}
\end{equation}
Then, one gets \cite{key-12}

\noindent 
\begin{equation}
\begin{array}{c}
\left(S_{total}\right)_{n-1}=4\pi\left(M^{2}-\frac{n-1}{2}\right)\\
\\
-\ln\left[4\pi\left(M^{2}-\frac{n-1}{2}\right)\right]+\frac{3}{32\pi\left(M^{2}-\frac{n-1}{2}\right)}
\end{array}\label{eq: entropia n-1}
\end{equation}

\noindent before the emission, and \cite{key-12} 
\begin{equation}
\begin{array}{c}
\left(S_{total}\right)_{n}=4\pi\left(M^{2}-\frac{n}{2}\right)\\
\\
-\ln\left[4\pi\left(M^{2}-\frac{n}{2}\right)\right]+\frac{3}{32\pi\left(M^{2}-\frac{n}{2}\right))}
\end{array}\label{eq: entropia n}
\end{equation}
after the emission, respectively. As a consequence, at level $n-1$
the BH has a number of micro-states 
\begin{equation}
\begin{array}{c}
g(N_{n-1})\propto\exp\{4\pi\left(M^{2}-\frac{n-1}{2}\right)+\\
\\
-\ln\left[4\pi\left(M^{2}-\frac{n-1}{2}\right)\right]+\frac{3}{32\pi\left(M^{2}-\frac{n-1}{2}\right))}\}
\end{array}\label{eq: microstati n-1}
\end{equation}
and, at level $n$, after the emission, the number of micro-states
is 
\begin{equation}
\begin{array}{c}
g(N_{n})\propto\exp\{4\pi\left(M^{2}-\frac{n}{2}\right)+\\
\\
-\ln\left[4\pi\left(M^{2}-\frac{n}{2}\right)\right]+\frac{3}{32\pi\left(M^{2}-\frac{n}{2}\right))}\}.
\end{array}\label{eq: microstati n}
\end{equation}
We note that when $n\;$ is large, but not enough large, it is also
$E_{n}\ll M_{n}\simeq M$ and one gets \cite{key-10,key-12}

\noindent 
\begin{equation}
\triangle A=32\pi M\Delta E_{n-1\rightarrow n},\label{eq: area quantum}
\end{equation}

\noindent 
\begin{equation}
N=\frac{A}{|\triangle A|}=\frac{16\pi M^{2}}{32\pi M\cdot\Delta E_{n-1\rightarrow n}}=\frac{M}{2\Delta E_{n-1\rightarrow n}},\label{eq: N}
\end{equation}

\noindent 
\begin{equation}
S_{BH}=\frac{A}{4}=8\pi NM\cdot\Delta E_{n-1\rightarrow n},\label{eq: Bekenstein-Hawking}
\end{equation}

\noindent 
\begin{equation}
S_{total}=8\pi NM\cdot\Delta E_{n-1\rightarrow n}-\ln\left[8\pi NM\cdot\Delta E_{n-1\rightarrow n}\right]+\frac{3}{64\pi NM\cdot\Delta E_{n-1\rightarrow n}},\label{eq: entropia totale 2}
\end{equation}

\noindent 
\begin{equation}
g(N)\propto\exp\left\{ 8\pi NM\cdot\Delta E_{n-1\rightarrow n}-\ln\left[8\pi NM\cdot\Delta E_{n-1\rightarrow n}\right]+\frac{3}{64\pi NM\cdot\Delta E_{n-1\rightarrow n}}\right\} .\label{eq: microstati}
\end{equation}
On the other hand, for large $n\;$ it is also $\Delta E_{n-1\rightarrow n}\approx\frac{1}{4M}$
\cite{key-10,key-12} and eqs. from (\ref{eq: area quantum}) to (\ref{eq: microstati})
become 
\begin{equation}
|\triangle A|\approx8\pi,\label{eq: Bekenstein}
\end{equation}

\begin{equation}
N\approx2M^{2},\label{eq: M square}
\end{equation}

\begin{equation}
S_{BH}\approx2\pi N,\label{eq: SBH approssimata}
\end{equation}

\noindent 
\begin{equation}
S_{total}\simeq2\pi N-\ln2\pi N+\frac{3}{16\pi N},\label{eq: entropia totale approssimata}
\end{equation}

\noindent 
\begin{equation}
g(N)\propto\exp\left[2\pi N-\ln\left(2\pi N\right)+\frac{3}{16\pi N}\right],\label{eq: microstati circa}
\end{equation}
which are are in agreement with previous literature \cite{key-25,key-35,key-36},
where the strictly thermal approximation has been used. 

Let us resume the way in which the BH model analysed in this Section
works. If $M$ is the original BH mass (in quantum terms the BH is
in its \emph{ground state}), after an high number of emissions (and
potential absorptions as the BH can capture neighboring particles),
the BH arrives at an excited level $n-1$ and its mass is now $M_{n-1}\equiv M-E_{n-1}$
where $E_{n-1}$ is the total energy emitted at that time and also
the absolute value of the frequency of the QNM associated to the excited
level $n-1$ \cite{key-12}. Now, the BH emits an energy $\Delta E_{n-1\rightarrow n}=E_{n}-E_{n-1}$
to jump to the subsequent level $n$. Thus, the BH mass decreases
again \cite{key-12} 
\begin{equation}
\begin{array}{c}
M_{n}\equiv M-E_{n-1}-\Delta E_{n-1\rightarrow n}=\\
\\
=M-E_{n-1}+E_{n-1}-E_{n}=M-E_{n}.
\end{array}\label{eq: masse}
\end{equation}
Notice that, in principle, the BH can bring back to the level $n-1$
absorbing an energy $-\Delta E_{n-1\rightarrow n}=\Delta E_{n\rightarrow n-1}=E_{n-1}-E_{n}$.
The quantum of area is \emph{the same} for both absorption and emission,
given by eq. (\ref{eq: 8 pi planck}), as one expects. 

One finds three different physical situations for excited BHs ($n\gg1$):
\cite{key-12}
\begin{enumerate}
\item $n\;$ is large, but not enough large as we have also $E_{n}\ll M_{n}\simeq M$
and we can use eqs. (\ref{eq: area quantum}), (\ref{eq: Bekenstein-Hawking}),
(\ref{eq: entropia totale 2}) and (\ref{eq: microstati}) which results
a better approximation than eqs. (\ref{eq: SBH approssimata}), (\ref{eq: entropia totale approssimata})
and (\ref{eq: microstati circa}) which were used in previous literature
in strictly thermal approximation, see \cite{key-25,key-35,key-36}
for example. This is indeed the approximation that we used in our
pioneering works \cite{key-10,key-11}.
\item $n\;$ is very much larger than in previous point, but the Planck
scale has not yet been approached. In that case, as it can be $E_{n}\lesssim M,$
$M_{n}\simeq M$ does not hold. One has to use the eqs. from (\ref{eq: Bekenstein-Hawking  n-1})
to (\ref{eq: microstati n}).
\item At the Planck scale $n\;$ is larger also than in previous point 2
and one needs a full quantum gravity theory, which is not yet known.
\end{enumerate}
We stress that, in our BH model, during a quantum jump a discrete
amount of energy is radiated and, for large values of the principal
quantum number $n,$ the analysis becomes independent from the other
quantum numbers. In a certain sense, QNMs represent the \textquotedbl{}electron\textquotedbl{}
which jumps from a level to another one and the absolute values of
the QNMs frequencies represent the energy \textquotedbl{}shells\textquotedbl{}.
In Bohr model \cite{key-43,key-44} electrons can only gain and lose
energy by jumping from one allowed energy shell to another, absorbing
or emitting radiation with an energy difference of the levels according
to the Planck relation (in standard units) $E=hf$, where $\: h\:$
is the Planck constant and $f\:$ the transition frequency. In our
BH model, QNMs can only gain and lose energy by jumping from one allowed
energy shell to another, absorbing or emitting radiation (emitted
radiation is given by Hawking quanta) with an energy difference of
the levels according to eq. (\ref{eq: variazione}). The similarity
is completed if one notes that the interpretation of eq. (\ref{eq: radice fisica})
is of a particle, the ``electron'', quantized on a circle of length
\cite{key-10,key-11,key-20} 
\begin{equation}
L=\frac{1}{T_{E}(E_{n})}=4\pi\left(M+\sqrt{M^{2}-\frac{n}{2}}\right),\label{eq: lunghezza cerchio}
\end{equation}
which is the analogous of the electron travelling in circular orbits
around the hydrogen nucleus, similar in structure to the solar system,
of Bohr model \cite{key-43,key-44}. On the other hand, Bohr model
is an approximated model of the hydrogen atom with respect to the
valence shell atom model of full quantum mechanics. In the same way,
the Bohr-like BH model should be an approximated model with respect
to the definitive, but at the present time unknown, BH model arising
from a full quantum gravity theory. 

Another key point is the following. In Hawking's original computation
\cite{key-1} if an emission can occur for a quantum of energy $E$,
then it can also occur for any other quantum of energy $bE$, where
$b$ is a continuous real parameter between $0$ and $\frac{M}{E},$
where $M$ is the BH mass. After the emission of a quantum of energy
$bE$, the BH radial coordinate is determined continuously by the
continuous parameter $b$. In other words, emissions of Hawking quanta
looks completely random. The situation looks to be similar within
the semi-classical context in which Parikh-Wilczek perform their calculation
\cite{key-2,key-3}. But here there is an important difference. The
discrete behavior in time of the radiation spectrum, in the sense
that we stressed in Section 2, implies the countable character of
the subsequent emitted Hawking quanta and, in turn, the correspondence
between the countable perturbations generated by the absorbed negative
energies and the BH QNMs. The fundamental consequence is that, differently
on Hawking's original computation \cite{key-1}, now emissions of
Hawking quanta are \emph{not} completely random. They are indeed governed
by eq. (\ref{eq: jump}). In fact, let us consider an emission from
the BH ground state to a state with large $n.$ After that, using
eq. (\ref{eq: n max}) (although we recall that the last area quantum
corresponds to the final Planck mass which is prevented to evaporate
by the Generalized Uncertainty Principle\cite{key-33}), one see that
the BH will have a finite and discrete number of potential emissions
given by 
\begin{equation}
n_{max}-n=2M^{2}-n.\label{eq: emissioni residue}
\end{equation}
It is enlightening to observe that such a number of potential residual
emissions, which is equal to the residual number of QNMs, is also
equal to the residual number of area quanta. In fact, by using eq.
(\ref{eq: me-1}) and recalling that $r_{H}=2M$ one easily compute
the area of the BH excited at level $n$ as 

\begin{equation}
A_{n}=16\pi M_{n}^{2}=16\pi\left(M^{2}-\frac{n}{2}\right),\label{eq: area n}
\end{equation}
which, dividing for the Bekenstein's area quantum $|\triangle A_{n}|=8\pi$
{[}30{]}, that we retrieved in eq. (\ref{eq: 8 pi planck}), gives
the number of area quanta for the BH excited at level $n$ 

\begin{equation}
N_{n}=2M^{2}-n.\label{eq: numeron quanti ad n}
\end{equation}
Thus, we understand that the key point is exactly Bekenstein's idea
on area quantization \cite{key-34}, i.e. as for large $n$ the BH
area is quantized, the BH can emit \emph{only} energies which are
consistent with such a quantization. In other words, emissions of
Hawking quanta are not completely random because the BH can emit only
energies which corresponds to reductions of its area which are multiples
of the Bekenstein's area quantum $|\triangle A_{n}|=8\pi$ given by
eq. (\ref{eq: 8 pi planck}). Hence, our results are completely consistent
with the idea that the Schwarzschild spacetime is quantized around
the BH core.

\section{Time evolution of Bohr-like black hole governed by a time dependent
Schrödinger equation: independent solution to the black hole information
paradox}

In his famous paper \cite{key-9} Hawking verbatim stated that ``\emph{Because
part of the information about the state of the system is lost down
the hole, the final situation is represented by a density matrix rather
than a pure quantum state}''. This statement was the starting point
of the popular \textquotedblleft{}\emph{BH information paradox}\textquotedblright{}. 

In Section 3, we naturally interpreted BH QNMs in terms of quantum
levels in a Bohr-like BH model. In this Section, following \cite{key-20},
we explicitly write down a \emph{time dependent Schrödinger equation
}for the system composed by Hawking radiation and BH QNMs. We show
that the physical state and the correspondent\emph{ wave function
}are written in terms of an \emph{unitary} evolution matrix instead
of a density matrix \cite{key-20}. As a consequence, the final state
results to be a \emph{pure} quantum state instead of mixed one \cite{key-20}.
Hence, Hawking's claim is falsified by the time evolution of the Bohr-like
BH model \cite{key-20}. BH evaporation results indeed to be an unitary
and time dependent process in which information comes out \cite{key-20}.
This is in full agreement with the assumption by 't Hooft that Schröedinger
equations can be used universally for all dynamics in the universe
\cite{key-37}. The final conclusion is that BH evaporation must be
information preserving \cite{key-20}.

In addition, it will be shown that the present approach permits also
to solve the entanglement problem connected with the information paradox
\cite{key-20}.

The BH information paradox is considered one of the most famous and
intriguing scientific controversies in the whole history of Science
\cite{key-20}. In classical general relativity, a BH is the ultimate
prison. Nothing can escape from it. As a consequence, when matter
falls into a BH, one can consider the information encoded as preserved
inside it, although inaccessible to outside observers. The celebrated
Hawking's discovery that quantum effects cause the BH to emit radiation
radically changed this situation \cite{key-1}. Hawking made a further
analysis \cite{key-9}, showing that the detailed form of the radiation
emitted by a BH should be thermal and independent of the structure
and composition of matter that collapsed to form the BH. In that way,
the radiation state results a completely mixed one which cannot carry
information on the BH's formation. After Hawking's original claim,
that we verbatim rewrote above, enormous time and effort have been
and are currently devoted to solve the information puzzle. In fact,
consequences of the BH information paradox are not trivial. As pure
quantum states arising from collapsed matter would decay into mixed
states if information is lost in BH evaporation, quantum gravity should
not be unitary \cite{key-38}! Various researchers worked and currently
work on the information puzzle. Some people think that quantum information
results destroyed in BH evaporation. Other people claim that the above
cited Hawking's statement was not correct and information is, instead,
preserved. A interesting and popular science book on the so called
``Black Hole War'' has been written by Susskind \cite{key-39} and
the paradox resulted introduced into physics folklore \cite{key-39,key-40}.
Two famous bets have been made by Hawking that BH does destroy information
\cite{key-39}. The first one, having Thorne like co-signer, with
Preskill, the latter with Page \cite{key-39}. In 2004-2005 Hawking
reversed his opinion claiming that information would probably be recovered
\cite{key-38,key-39}. Various attempts to solve the information puzzle
and historical notes on the controversy can be found in \cite{key-37}-\cite{key-41}.
Recently, Hawking reversed his opinion again, with a couples of ambiguous
statements verbatim claiming that ``\emph{The chaotic collapsed object
will radiate deterministically but chaotically. It will be like weather
forecasting on Earth. That is unitary, but chaotic, so there is effective
information loss. One can't predict the weather more than a few days
in advance}'' \cite{key-48}.

As we previously recalled, a key point, concerning not only the BH
information paradox, but the whole BH quantum physics, is that the
BH radiation spectrum is not strictly thermal \cite{key-2,key-3},
differently from Hawking's original computations \cite{key-1,key-9}.
Now, we show that the time evolution of the Bohr-like BH model is
governed by a time dependent Schrödinger equation which enables pure
quantum states to evolve to pure quantum states in a unitary evolution
which preserves quantum information and, in turn, falsifies the above
cited statement by Hawking \cite{key-20}. It will be also shown that,
in addition, the following approach solves the entanglement problem
connected with the information paradox \cite{key-20}.

Let us start by recalling that $E_{n}\:$ in eq. (\ref{eq: radice fisica})
is interpreted like the total energy emitted by the BH at that time,
i.e. when the BH is excited at a level $n$, see Section 3 and refs.
\cite{key-12,key-20}. If one considers an emission from the ground
state to a state with large $n\:$ and uses eq. (\ref{eq: radice fisica}),
the BH mass changes from $M\:$ to \cite{key-20}

\begin{equation}
M_{n}\equiv M-E_{n}=\sqrt{M^{2}-\frac{n}{2}}.\label{eq: mn}
\end{equation}
In the transition from the state with $n$ to a state with $m>n$
the BH mass changes again from $M_{n}\:$ to \cite{key-20}

\begin{equation}
\begin{array}{c}
M_{m}\equiv M_{n}-\Delta E_{n\rightarrow m}=M-E_{m}\\
=\sqrt{M^{2}-\frac{m}{2}},
\end{array}\label{eq: mm}
\end{equation}
where $\Delta E_{n\rightarrow m}\equiv E_{m}-E_{n}=M_{n}-M_{m}$ is
the jump between the two levels due to the emission of a particle
having frequency $\omega_{n,m}=\Delta E_{n\rightarrow m}$ \cite{key-20}.
Let us show that the energy emitted in an arbitrary transition from
$n$ to $m$, where $m>n$ (we are considering an emission), is proportional
to the effective temperature $\left[T_{E}\right]_{n\rightarrow m}$
associated to the transition \cite{key-20}. Putting \cite{key-20}
\begin{equation}
\Delta E_{n\rightarrow m}\equiv E_{m}-E_{n}=M_{n}-M_{m}=K\left[T_{E}\right]_{n\rightarrow m},\label{eq: differenza radici fisiche}
\end{equation}
where $M_{n}$ and $M_{m}$ are given by eqs. (\ref{eq: mn}) and
(\ref{eq: mm}) , we search values of the constant $K$ for which
eq. (\ref{eq: differenza radici fisiche}) is satisfied. As discussed
in \cite{key-8}, \cite{key-10}-\cite{key-14}, \cite{key-20,key-46}
and in Section 2, the effective temperature is the inverse of the
average value of the inverses of the initial and final Hawking temperatures

\begin{equation}
\left[T_{E}\right]_{n\rightarrow m}=\frac{1}{4\pi\left(M_{n}+M_{m}\right)}.\label{eq: temperatura efficace di transizione}
\end{equation}
Hence, one rewrites eq. (\ref{eq: differenza radici fisiche}) as
\cite{key-20}

\begin{equation}
\Delta E_{n\rightarrow m}=M_{n}^{2}-M_{m}^{2}=\frac{K}{4\pi}.\label{eq: differenza radici fisiche 2}
\end{equation}
By using eqs. (\ref{eq: mn}) and (\ref{eq: mm}) eq. (\ref{eq: differenza radici fisiche 2})
becomes \cite{key-20}

\begin{equation}
\frac{1}{2}\left(m-n\right)=\frac{K}{4\pi}.\label{eq: K solved}
\end{equation}
Thus, eq. (\ref{eq: differenza radici fisiche}) is satisfied for
$K=2\pi\left(m-n\right),$ and we find \cite{key-20} 
\begin{equation}
\Delta E_{n\rightarrow m}=2\pi\left(m-n\right)\left[T_{E}\right]_{n\rightarrow m}.\label{eq: differenza radici fisiche finale}
\end{equation}
Considering eq. (\ref{eq: Corda Probability}), we can write the probability
of emission between the two levels $n$ and $m\;$ in an elegant form
\cite{key-20} 
\begin{equation}
\Gamma_{n\rightarrow m}=\alpha\exp-\left\{ \frac{\Delta E_{n\rightarrow m}}{\left[T_{E}\right]_{n\rightarrow m}}\right\} =\alpha\exp\left[-2\pi\left(m-n\right)\right],\label{eq: Corda Probability Intriguing}
\end{equation}
where the pre-factor is $\alpha\sim1.$ Then, one finds that the probability
of emission between two arbitrary levels $n$ and $m$ is proportional
to $\exp\left[-2\pi\left(m-n\right)\right].$ We observe that the
probability of emission has its maximum value $\sim\exp(-2\pi)$ for
$m=n+1$ i.e. for two adjacent levels, as we intuitively expects \cite{key-20}. 

In a quantum mechanical framework, we physically interpret emissions
of Hawking quanta like quantum jumps among the unperturbed levels
(\ref{eq: radice fisica}) \cite{key-10,key-11,key-12,key-20}. Following
\cite{key-20,key-42}, the time evolution of perturbations can be
described by the operator 

\emph{
\begin{equation}
U(t)=\begin{array}{c}
W(t)\;\;\; for\;0\leq t\leq\tau\\
0\;\;\; for\; t<0\; and\; t>\tau,
\end{array}\label{eq: perturbazione}
\end{equation}
}and the complete (time dependent) Hamiltonian is described by the
operator \cite{key-20,key-42}

\begin{equation}
H(x,t)\equiv V(x)+U(t),\label{eq: Hamiltoniana completa}
\end{equation}
where $V(x)$ is the \emph{effective Regge-Wheeler} potential (\ref{eq: effettiva 1})
of the time independent \emph{effective Schröedinger-like equation}
(\ref{eq: diff.-1}). Then, considering a wave function $\psi(x,t),$
we can write the correspondent \emph{time dependent Schroedinger equation
}for the system as \cite{key-20,key-42}

\begin{equation}
i\frac{d|\psi(x,t)>}{dt}=\left[V(x)+U(t)\right]|\psi(x,t)>=H(x,t)|\psi(x,t)>.\label{eq: Schroedinger equation}
\end{equation}
If $\varphi_{m}(x)$ and $\omega_{m}$ are the eigenfunctions of the
time independent Schröedinger-like equation (\ref{eq: diff.-1}) and
the correspondent eigenvalues respectively, the\emph{ }state satisfying
eq. (\ref{eq: Schroedinger equation}) is \cite{key-20,key-42}

\begin{equation}
|\psi(x,t)>=\sum_{m}a_{m}(t)\exp\left(-i\omega_{m}t\right)|\varphi_{m}(x)>.\label{eq: Schroedinger wave-function}
\end{equation}
We consider Dirac delta perturbations \cite{key-10,key-11,key-12,key-20,key-25}
which represent subsequent absorptions of particles having negative
energies being associated to emissions of Hawking quanta in the mechanism
of particle pair creation. In the basis $|\varphi_{m}(x)>$, the matrix
elements of $W(t)$ are \cite{key-20,key-42}

\begin{equation}
W_{ij}(t)\equiv A_{ij}\delta(t),\label{eq: a delta}
\end{equation}
with $W_{ij}(t)=<\varphi_{i}(x)|W(t)|\varphi_{j}(x)>$ \cite{key-20,key-42}
and the $A_{ij}$ are real. As we want to solve the complete quantum
mechanical problem described by the operator (\ref{eq: Hamiltoniana completa}),
we need to know the probability amplitudes $a_{m}(t)$ due to the
application of the perturbation described by the time dependent operator
(\ref{eq: perturbazione}) \cite{key-20,key-42}, representing the
perturbation associated to the emission of a Hawking quantum \cite{key-20,key-42}.
For $t<0,$ i.e. before the perturbation operator (\ref{eq: perturbazione})
starts to work, the system is in a stationary state $|\varphi_{n}(t,x)>,$
at the quantum level $n,$ with energy $E_{n}=|\omega_{n}|$ given
by eq. (\ref{eq: radice fisica}) \cite{key-20,key-42}. Thus, only
the term

\begin{equation}
|\psi_{n}(x,t)>=\exp\left(-i\omega_{n}t\right)|\varphi_{n}(x)>,\label{eq: Schroedinger wave-function in.}
\end{equation}
in eq. (\ref{eq: Schroedinger wave-function}) is not null for $t<0.$
This implies $a_{m}(t)=\delta_{mn}\:\:$for $\: t<0.$ After the emission
the perturbation operator (\ref{eq: perturbazione}) stops to work
and for $t>\tau$ the probability amplitudes $a_{m}(t)$ return to
be time independent, having the value $a_{n\rightarrow m}(\tau)$
\cite{key-20,key-42}. In other words, for $t>\tau\:$ the system
is in the state \cite{key-20,key-42}

\begin{equation}
|\psi_{final}(x,t)>=\sum_{m=n}^{m_{max}}a_{n\rightarrow m}(\tau)\exp\left(-i\omega_{m}t\right)|\varphi_{m}(x)>,\label{eq: Schroedinger wave-function fin.}
\end{equation}
described by the\emph{ wave function $\psi_{final}(x,t)$ }\cite{key-20,key-42},
and one sees that the probability to find the system in an eigenstate
having energy $E_{m}=|\omega_{m}|$ is \cite{key-20,key-42}

\begin{equation}
\Gamma_{n\rightarrow m}(\tau)=|a_{n\rightarrow m}(\tau)|^{2}.\label{eq: ampiezza e probability}
\end{equation}
A standard analysis will give the following differential equation
from eq. (\ref{eq: Schroedinger wave-function fin.}) \cite{key-20,key-42}

\begin{equation}
i\frac{d}{dt}a_{n\rightarrow m}(t)=\sum_{l=m}^{m_{max}}W_{ml}a_{n\rightarrow l}(t)\exp\left[i\left(\Delta E_{l\rightarrow m}\right)t\right].\label{eq: systema differenziale}
\end{equation}
The Dayson series permits to obtain the solution \cite{key-20,key-42}

\begin{equation}
a_{n\rightarrow m}=-i\int_{0}^{t}\left\{ W_{mn}(t')\exp\left[i\left(\Delta E_{n\rightarrow m}\right)t'\right]\right\} dt',\label{eq: solution}
\end{equation}
to first order in $U(t)$. Inserting (\ref{eq: a delta}) in (\ref{eq: solution})
we get \cite{key-20,key-42} 
\begin{equation}
a_{n\rightarrow m}=iA_{mn}\int_{0}^{t}\left\{ \delta(t')\exp\left[i\left(\Delta E_{n\rightarrow m}\right)t'\right]\right\} dt'=\frac{i}{2}A_{mn}.\label{eq: solution 2}
\end{equation}
Combining eq. (\ref{eq: solution 2}) with eqs. (\ref{eq: Corda Probability Intriguing})
and (\ref{eq: ampiezza e probability}) at the end we obtain \cite{key-20,key-42}

\begin{equation}
\begin{array}{c}
\alpha\exp\left[-2\pi\left(m-n\right)\right]=\frac{1}{4}A_{mn}^{2}\\
\\
A_{mn}=2\sqrt{\alpha}\exp\left[-\pi\left(m-n\right)\right]\\
\\
a_{n\rightarrow m}=-i\sqrt{\alpha}\exp\left[-\pi\left(m-n\right)\right].
\end{array}\label{eq: uguale}
\end{equation}
We recall that it is $\sqrt{\alpha}\sim1.$ Then we get $A_{mn}\sim10^{-2}$
for $m=n+1$, i.e. when the probability of emission has its maximum
value \cite{key-20,key-42}. Therefore, second order terms in $U(t)$
are $\sim10^{-4},$ which means that our approximate result to first
order in $U(t)$ is very good \cite{key-20,key-42}. We note that
for $m>n+1$ the approximation is better because the $A_{mn}$ are
even smaller than $10^{-2}$. Then, the final form of the ket representing
the state is \cite{key-20,key-42}

\begin{equation}
|\psi_{final}(x,t)>=\sum_{m=n}^{m_{max}}-i\sqrt{\alpha}\exp\left[-\pi\left(m-n\right)-i\omega_{m}t\right]|\varphi_{m}(x)>.\label{eq: Schroedinger wave-function finalissima}
\end{equation}
The\emph{ }state (\ref{eq: Schroedinger wave-function finalissima})
represents a \emph{pure final state instead of a mixed final state}
and the states are written in terms of an \emph{unitary} evolution
matrix instead of a density matrix \cite{key-20,key-42}.\emph{ }Therefore,
one finds that\emph{ }information is not loss in BH evaporation \cite{key-20,key-42}.
The result is in full agreement with the assumption by 't Hooft that
Schrödinger equations can be used universally for all dynamics in
the universe \cite{key-37}. 

We observe that the final state of eq. (\ref{eq: Schroedinger wave-function finalissima})
is due to potential emissions of Hawking quanta having negative energies
which perturb the BH and ``trigger'' the QNMs corresponding to potential
arbitrary transitions $n\rightarrow m$, with $m>n$ \cite{key-20}.
Then, the subsequent \emph{collapse of the wave function} to a new
a stationary state \cite{key-20} 
\begin{equation}
|\psi_{m}(x,t)>=\exp\left(-i\omega_{m}t\right)|\varphi_{m}(x)>,\label{eq: Schroedinger wave-function out}
\end{equation}
at the quantum level $m,$ implies that the wave function of the particle
having negative energy $-\Delta E_{n\rightarrow m}=\omega_{n}-\omega_{m}$
has been transferred to the QNM and it is given by \cite{key-20}
\begin{equation}
|\psi_{-\left(m-n\right)}(x,t)>\equiv-\exp\left[i(\omega_{m}-\omega_{n})t\right]\left[|\varphi_{m}(x)>-|\varphi_{n}(x)>\right].\label{eq: funzione onda particella emessa}
\end{equation}
The wave function (\ref{eq: funzione onda particella emessa}) results
entangled with the wave function of the particle with positive energy
propagating towards infinity in the mechanism of particle creation
by BHs. Below it will be shown that this key point solves the entanglement
problem connected with the information paradox \cite{key-20}.

Our analysis is strictly correct only for excited BHs, i.e. for $n\gg1$
\cite{key-20,key-42}. For this reason we assumed an emission from
the ground state to a state with large $n\:$ in the discussion \cite{key-20,key-42}.
On the other hand, as we have seen in Section 4, a state with large
$n\:$ is always reached at late times, maybe not through a sole emission
from the ground state, but through various subsequent emissions and
potential absorptions \cite{key-20,key-42}. 

Now, let us discuss another key point, which concerns quantum entanglement.
We could think that, although previous analysis discusses a very natural
model of Hawking radiation and BH evaporation, there is no reference
to the BH spacetime, where information is assumed to be conserved.
There are indeed authors who claim that the real challenge in solving
the information paradox is to reconcile models of Hawking radiation
with the spacetime structure within the BH horizon, where the quantum
information falling into the singularity is causally separated from
the outgoing Hawking quanta, see the work by Mathur \cite{key-41}
for example. In any case, this kind of criticism does not work for
the analysis in this Section. In the above analysis there is indeed
a subtle connection between the emitting Hawking quanta and the BH
spacetime within the horizon, where information is conserved. This
approach to the BH information problem concerns the entanglement structure
of the wave function which is associated to the particle pair creation
\cite{key-41}. In fact, in order to solve the information puzzle,
we need to know the part of the wave function in the interior of the
BH horizon \cite{key-41}, i.e. the part of the wave function associated
to the particle having negative energy in the tunnelling mechanism.
In the emissions of Hawking quanta, this is exactly the part of the
wave function which results entangled with the part of the wave function
outside, i.e. the part of the wave function associated to the particle
which has positive energy and escapes from the BH \cite{key-41}.
If we ignore such an interior part of the wave function, we miss the
entanglement completely, failing to understand the information problem
\cite{key-41}. But when one considers the above discussed correspondence
between Hawking radiation and BH QNMs, the particle which has negative
energy and falls into the singularity transfers its part of the wave
function and, in turn, the information encoded in such a part of the
wave function, to the QNM. In other terms, the emitted quanta are
entangled with BH QNMs, which are the oscillations of the BH horizon.
This key point is exactly the subtle connection between the emitted
Hawking quanta and the BH spacetime that we need to find. We explain
this important issue in detail. Again, we emphasize that the correspondence
between emitted radiation and proper oscillation of the emitting body
is a fundamental behavior of every radiation process in Nature, and
this issue helps to solve the entanglement problem. The mechanism
of particles creation by BHs \cite{key-1} has been described as tunnelling
arising from vacuum fluctuations near the BH horizon in \cite{key-2}-\cite{key-8}
and in the Introduction of this paper. Let us again assume an initial
emission from the BH ground state to a state having large $n,$ say
$n=n_{1}\gg1.$ The absorbed particle, which has negative energy $-|\omega_{n_{1}}|$,
generates a QNM which has an energy-frequency $E_{n_{1}}=|\omega_{n_{1}}|$.
As a consequence, the BH mass changes from $M\:$ to 

\begin{equation}
M_{n_{1}}\equiv M-E_{n_{1}}=\sqrt{M^{2}-\frac{n_{1}}{2}}.\label{eq: m1}
\end{equation}
Thus, the energy of the first particle absorbed by the BH, which has
negative energy is transferred, together with its part of the wave
function, to the QNM which is, in turn, entangled with the emitted
particle having positive energy. Let us consider eq. (\ref{eq: funzione onda particella emessa}).
If one sets $n=0$ and $m=n_{1}$ one finds that the part of the wave
function in the interior of the horizon, i.e. the part of the wave
function associated to the particle having negative energy (infalling
mode) which has been transferred to the QNM is \cite{key-20} 
\begin{equation}
|\psi_{-n_{1}}(x,t)>=-\exp\left(i\omega_{n_{1}}t\right)|\varphi_{n_{1}}(x)>.\label{eq: da zero ad n1}
\end{equation}

Now, we consider a second emission. This new emission corresponds
to the transition from the state with $n=n_{1}$ to another state
with, say, $n=n_{2}>n_{1}$. The BH mass changes from $M_{n_{1}}\:$
to

\begin{equation}
\begin{array}{c}
M_{n_{2}}\equiv M_{n_{1}}-\Delta E_{n_{1}\rightarrow n_{2}}=M-E_{n_{2}}\\
=\sqrt{M^{2}-\frac{n_{2}}{2}},
\end{array}\label{eq: m2}
\end{equation}
and $\Delta E_{n_{1}\rightarrow n_{2}}\equiv E_{n_{2}}-E_{n_{1}}=M_{n_{1}}-M_{n_{2}}$
is the jump between the two levels. The energy of the second particle
absorbed by the BH which has negative energy is transferred, together
with its part of the wave function, again to the QNM, which has an
increased energy-frequency $E_{n_{2}}=|\omega_{n_{12}}|$ and is now
entangled with both the two emitted particles which have positive
energy. If one uses again eq. (\ref{eq: funzione onda particella emessa})
and sets $n=n_{1}$ and $m=n_{2}$, one finds that the part of the
wave function of the second infalling mode which has been transferred
to the QNM is \cite{key-20} 
\begin{equation}
|\psi_{-\left(n_{2}-n_{1}\right)}(x,t)>=-\exp\left[i(\omega_{n_{2}}-\omega_{n_{1}})t\right]\left[|\varphi_{n_{2}}(x)>-|\varphi_{n_{1}}(x)>\right].\label{eq: da n1 ad n2}
\end{equation}

Let us consider a third emission, corresponding to the transition
from the state with $n=n_{2}$ to a further different state with,
say, $n=n_{3}>n_{2}$. Now, the BH mass changes from $M_{n_{2}}\:$
to

\begin{equation}
\begin{array}{c}
M_{n_{3}}\equiv M_{n_{2}}-\Delta E_{n_{2}\rightarrow n_{3}}=M-E_{n_{3}}\\
=\sqrt{M^{2}-\frac{n_{3}}{2}},
\end{array}\label{eq: m3}
\end{equation}
where $\Delta E_{n_{2}\rightarrow n_{3}}\equiv E_{n_{3}}-E_{n_{2}}=M_{n_{2}}-M_{n_{3}}$
is the jump between the two levels. Again, the energy of the third
particle absorbed by the BH and having negative energy is transferred,
together with its part of the wave function, to the QNM which has
now a further increased energy-frequency $E_{n_{13}}=|\omega_{n_{3}}|$
and is entangled with the three emitted particles which have positive
energy. Now, eq. (\ref{eq: funzione onda particella emessa}) with
$n=n_{2}$ and $m=n_{3}$ gives the part of the wave function of the
third infalling mode which has been transferred to the QNM as 
\begin{equation}
|\psi_{-\left(n_{3}-n_{2}\right)}(x,t)>=-\exp\left[i(\omega_{n_{3}}-\omega_{n_{2}})t\right]\left[|\varphi_{n_{3}}(x)>-|\varphi_{n_{2}}(x)>\right].\label{eq: da n2 ad n3}
\end{equation}

The process will continue again, and again, and again... till the
\emph{Planck distance} and the \emph{Planck mass} are approached by
the evaporating BH. At that point, the Generalized Uncertainty Principle
prevents the total BH evaporation, see Section 3 and \cite{key-12,key-33}
, and we need a full theory of quantum gravity for the further evolution. 

In any case, we emphasize again that the energy $E_{n}\:$ of the
generic QNM having principal quantum number $n$ is interpreted like
the total energy emitted by the BH at that time, i.e. when the BH
is excited at a level $n$ \cite{key-12,key-20}. As a consequence,
such a QNM is entangled with all the Hawking quanta emitted at that
time.

Therefore, all the quantum physical information which is fallen into
the singularity is not causally separated from the outgoing Hawking
radiation, but is instead recovered and codified in eq. (\ref{eq: Schroedinger wave-function finalissima})
through the correspondence between Hawking radiation and BH QNMs.
Following Mathur \cite{key-41}, the solution to the information puzzle
is to find a physical effect that we could have have missed. In this
Section, we have shown that the natural correspondence between Hawking
radiation and BH QNMs which governs the BH evaporation \emph{is exactly
that missed physical effect}.

\section{Conclusion remarks}

In this review paper some recent important results in BH quantum physics,
which concern the BH effective state and the Bohr-like model for BHs
in \cite{key-10,key-11,key-12,key-20} have been reanalyzed. The correspondence
between Hawking radiation and BH QNMs permits indeed to naturally
consider QNMs as BH quantum levels in a semi-classical model somewhat
similar to the historical semi-classical model of the structure of
a hydrogen atom introduced by Bohr in 1913 \cite{key-43,key-44,key-45}.
In the Bohr-like BH in a certain sense QNMs represent the \textquotedbl{}electron\textquotedbl{}
jumping from a level to another one and the absolute values of the
QNMs frequencies ``triggered'' by emissions (Hawking radiation)
and absorption of particles represent the energy \textquotedbl{}shells\textquotedbl{}
of the ``gravitational hydrogen atom''. 

Again, we stress that Bohr model is an approximated model of the hydrogen
atom with respect to the valence shell atom model of full quantum
mechanics. Then, one expects the Bohr-like BH model to be an approximated
model with respect to the definitive, but at the present time unknown,
BH model arising from a complete theory of quantum gravity. 

Important consequences on the BH information puzzle have been also
discussed, reviewing the independent solution to the paradox found
in \cite{key-20}. The system Hawking radiation - BH QNMs obeys indeed
to a time dependent Schrödinger equation which permits the final BH
state to be a \emph{pure} quantum state instead of mixed one in perfect
agreement with the assumption by 't Hooft that Schröedinger equations
can be used universally for all dynamics in the universe \cite{key-37}.
We have also shown that our approach also solves the entanglement
problem connected with the information paradox, a issue raised in
\cite{key-41}. 

Finally, for the sake of completeness \cite{key-49}, we recall that,
in some cases in extended gravity \cite{key-50,key-51,key-52} the
semi-classical effects may lead to instabilities of BHs, with strange
effects like anti-evaporation. These instabilities may qualitatively
change QNMs, or even make their emergence impossible \cite{key-49}.

\section{Acknowledgements }

It is a pleasure to thank Prof. Maxim Khlopov for inviting me to write
this review paper. I thank the unknown Reviewers for useful comments.
I thank Prof. J. D. Bekenstein for pointing out to me some important
issues concerning the history of BH thermodynamics.

\section{Conflict of Interests}

The author declares that there is no conflict of interests regarding
the publication of this article.

\end{document}